\documentclass[rmp,showpacs,twocolumn]{revtex4}

\usepackage[english]{babel}
\usepackage{amssymb}
\usepackage{amsfonts}
\usepackage{amsmath}

\usepackage{graphicx}

\begin{document}
\title{Translocation of a Single Stranded DNA Through \\
 a Conformationally Changing Nanopore}

\author{O. Flomenbom and J. Klafter}
\address{\it School of Chemistry, Raymond \& Beverly Sackler Faculty of Exact Sciences,\\  Tel-Aviv University, \it
Tel-Aviv 69978, Israel}

\date{\today}

\begin{abstract}
We investigate the translocation of a single stranded DNA through
a pore which fluctuates between two conformations, using coupled
master equations. The probability density function of the first
passage times (FPT) of the translocation process is calculated,
displaying a triple, double or mono peaked behavior, depending on
the interconversion rates between the conformations, the applied
electric field, and the initial conditions. The cumulative
probability function of the FPT, in a field-free environment, is
shown to have two regimes, characterized by fast and slow
timescales. An analytical expression for the mean first passage
time of the translocation process is derived, and provides, in
addition to the interconversion rates, an extensive
characterization of the translocation process. Relationships to
experimental observations are discussed.
\end{abstract}

\pacs{ 87.14.Gg, 87.15.Aa,87.15.He}

\maketitle

\begin{flushleft}
{\bf INTRODUCTION}\end{flushleft}

Translocation of biopolymers through a membrane pore occurs in a
variety of biological processes, such as gene expression in
eucaryotic cells \cite{cell}, conjugation between procaryotic
cells, and virus infection \cite{micro}. The importance of
translocation in biological systems and its applications, have
been the motivation for recent theoretical and experimental work
on this topic. In experiments one usually measures the time it
takes a single voltage-driven single stranded DNA (ssDNA) to
translocate through $\alpha$-$hemolysin$ channel of a known
structure \cite{song}. Since ssDNA is negatively charged (each
monomer of length $b$ has an effective charge of $zq$, where $q$
is the electron charge, and $z$ ($0<z<1$ ) is controlled by the
solution pH and strength), when applying a voltage the polymer is
subject to a driving force while passing through the transmembrane
pore part ({\it TPP}) from the negative (cis) side to the positive
(trans) side (for an illustration of the process see Meller 2003,
figure 4). Because the presence of the ssDNA in the {\it TPP}
blocks the cross-{\it TPP} current, one can deduce the FPT
probability density function ({\it pdf}), $F(t)$, from the current
blockade duration times \cite{kasa1,meller1}.

Experiments by Kasianowicz et al. (1996), show $F(t)$ with three
peaks. It was suggested that the short-time peak represents the
non-translocated events, while the other two, longer-time peaks,
represent translocation events of different ssDNA orientations. In
addition, the times that maximize the translocation peaks were
shown to be proportional to the polymer length and inversely
proportional to the applied field. In experiments by Meller et al.
(2001), $F(t)$ was shown to be mono-peaked, with a corresponding
maximizing time that has an inverse quadratic field dependence.
More recently, Bates et al. (2003) measured the FPT cumulative
probability function ({\it cpf}), which is the probability to exit
the channel until time $t$, $G(t)=\int_0^tF(s)ds$, in a field-free
environment. $G(t)$ was approximated by two well separated
timescales with the ratio of $1/20$.

In previous theoretical works, the translocation of a ssDNA
through a nanopore was described by statistical models that
focused on calculating the free energy of the process as a
function of the translocation state. The free energy contained
terms that represent the entropy and the chemical potential of the
polymer parts on both sides of a zero thickness membrane (Sung and
park, 1996; Muthukumar, 1999). The role of the membrane thickness
was studied by Ambjornsson et al. (2002), Solonika and Kolomeisky
(2003), and Flomenbom and Klafter (2003a). The obtained free
energy was used to calculate the mean first passage time (MFPT),
which asymptotically was found to scale linearly with the polymer
length for a field-biased process. This is the expected MFPT
dependence of a Markovian biased random walk in a finite interval
\cite{redner}.

A different approach was suggested by Lubensky and Nelson (1999)
and was further developed by Brezhkovskii and Gopich (2003), where
a diffusion-convection equation was used to describe the
translocation process, under the assumption that the polymer parts
outside the membrane hardly affect the translocation. Brezhkovskii
and Gopich (2003) showed that by changing the cis absorbing end to
be partially absorbing, the mono-peaked $F(t)$ obtained by
Lubensky and Nelson (1999) changes to a superposition of a
decaying non-translocation {\it pdf} and a peaked translocation
{\it pdf}. Using the fractional Fokker-Planck equation, Metzler
and Klafter (2003) suggested an explanation for the slow
relaxation time of the experimentally obtained $G(t)$. We have
shown by using the master equation (ME) that $F(t)$ can be double
or mono peaked, depending on the applied field and on the initial
condition \cite{flom}.

In the approaches summarized above the structure of the pore is
taken to be rigid, namely, governed by a single conformation.
Although it is known that the $\alpha$-$hemolysin$ channel has a
rigid structure that allows its crystallization \cite{song},
during the translocation of a long polymer (larger than the pore
length) with the same width that of the channel at some cross
sections along the channel, small fluctuation in the channel
structure may occur, which give rise to a more complex process
than what was assumed so far. In this work we relax the assumption
of a single pore conformation and introduce a second conformation
coupled to the first one. In a continuum formalism, the process
takes place in an effective two dimensional system, where one
dimension represents the translocation itself, and the second
dimension represents the structural fluctuations. This picture is
richer and is more realistic, since small structural changes in
physiological conditions are known to occur in large biomolecules,
certainly during interaction with other large biomolecules.

The function that represents best the translocation process is
$F(t)$ (or its integral $G(t)$). Through the dependence of $F(t)$
on the system parameters, we learn about the important degrees of
freedom which participates in the translocation process. The
characteristics of $F(t)$ are the dependence of its shape,
moments, and times that maximize its peaks on the system
parameters. Using the generalized model that takes into account
fluctuations in the pore structure, we calculate $F(t)$ and show
that it can display one two or three peaks, depending on the
applied voltage, the temperature and the interconversion rates
between the two conformations. Analytical expressions for the MFPT
are derived and related to the experimental findings. In addition,
we calculate the cumulative probability $G(t)$ in the field-free
limit, and show that it also provides valuable information about
the system parameters. Thus, these tools help in gaining insight
into the translocation of a polymer through a narrow pore, and in
explaining the diversity of the experimental observations
\cite{kasa1,meller1}.

\begin{flushleft}
{\bf THEORETICAL MODELLING}
\end{flushleft}
\begin{flushleft}
{\bf {Basic Model}}
\end{flushleft}

The basic model we use to describe the translocation relies on a
one-dimensional process. To use this simplification, we map the
three-dimensional translocation process onto a discrete
one-dimensional space containing $n(=N+d-1)$ states separated from
each other by a unit length $b$. The translocation takes place
within a {\it TPP} of a length that corresponds to $d$ monomers.
An $n$-state ME is introduced to describe the translocation of an
$N$-monomer long ssDNA subject to an external voltage $V$ and
temperature $T$. The occupation $pdf$ of the $j$ state is
$[\overrightarrow{P}(t)]_j$=$P_j(t)$, where the state index $j$
determines the number of monomers on each side of the membrane and
within the {\it TPP} ($m_j$). $P_j(t)$ satisfies the equation of
motion:
\begin{eqnarray}\label{1}
\partial P_j(t)/\partial
t=a_{j+1,j}P_{j+1}(t)+a_{j-1,j}P_{j-1}(t)-\nonumber\\
-(a_{j,j+1}+a_{j,j-1})P_j(t),~~~~~~~~~~~~~
\end{eqnarray}
under absorbing boundary conditions on both sides of the membrane
(the polymer can exit the {\it TPP} on both sides). Eq. 1 can be
written in a matrix representation:
\begin{equation}\label{2}
 \partial
\overrightarrow{P}(t)/\partial t = \mathbf{A}
\overrightarrow{P}(t),
\end{equation}
where the propagation matrix $\mathbf{A}$ is a tridiagonal matrix
that contains information about the transitions between states in
terms of rate constants, $a_{j,j\pm1}$, which are given by:
\begin{equation}\label{3}
a_{j,j\pm1}=k_{j}p_{j,j\pm1}.
 \end{equation}
Here $k_j$ is the rate to perform a step, $p_{j,j-1}$
($p_{j,j+1}$) is the probability to move one state from state $j$
to the trans (cis) side, and $p_{j,j+1}+p_{j,j-1}=1$.

$k_j$ is taken to be similar to the longest bulk relaxation time
of a polymer \cite{doi}:
\begin{equation}\label{4}
 k_j=1/(\beta
\xi_{p}b^2m_{j}^\mu)\equiv R/m_{j}^\mu;~~~\beta^{-1}\equiv k_BT,
\end{equation}
with two exceptions: the parameter $\xi_{p}$ represents the
ssDNA-{\it TPP} interaction and cannot be calculated from the
Stokes relation, and $\mu$ serves as a measure of the polymer
stiffness inside the confined volume of the {\it TPP}, and is
bounded by the conventional values \cite{doi}: $0\leq\mu\leq 1.5$.

Assuming a quasi-equilibrium process, which enables using the
detailed balance condition, and using the approximation
$a_{j,j-1}/a_{j-1,j}\approx p_{j,j-1}/(1-p_{j,j-1})$, the
probability $p_{j,j-1}$ is found to be:
\begin{equation}\label{5}
  p_{j,j-1}=(1+e^{\beta\Delta E_j})^{-1}.
\end{equation}
The free energy difference between states, $\Delta
E_j=E_{j-1}-E_j$, is computed considering three contributions:
electrostatic, entropic, and an average attractive interaction
energy between the ssDNA and the pore. More explicitly,
$\beta\Delta E_j$ is given by $\beta\Delta E_j=\beta\Delta
E_j^p+\delta_j$, where $\beta\Delta E_j^p \leq 0$ represents the
effect of the field which directs towards the trans-side and
$\delta_j>0$ (for $j>d$) represents an effective directionality to
the cis-side, which originates from the entropic factors and the
average attractive interaction energy between the ssDNA and the
pore. For a more detailed discussion see Flomenbom and Klafter
(2003a).

Several features emerge from the simple one-dimensional model. For
homopolymers, poly-$d nu$, where $nu$ stands for the nucleotide
type, we estimate $\xi_p(A_{nu})\approx 10^{-4}meVs/nm^2$,
$\xi_p(C_{nu})$=$\xi_p(T_{nu})$=$\xi_p(A_{nu})/3$ and
$\mu(C_{nu})$=$1$, $\mu(A_{nu})$=$1.14$, $\mu(T_{nu})$=$1.28$.
Here $A_{nu}$, $C_{nu}$ and $T_{nu}$ stand for adenine, cytosine
and thymine nucleotides, respectively. Interestingly, $\xi_p$ is
three order of magnitude larger than the bulk friction constant,
which is consistent with the role assign to $\xi_p$ to represent
the interaction between the polymer and the channel.

In addition, from the expressions for $\beta \Delta E_j$ and
$p_{j,j-1}$, the important parameter $V/V_C\equiv \beta
z|q|V(1+1/d)$ comes out naturally. This ratio determines the
directionality of the translocation, and, in particular, for
$V/V_C>1$ there is a bias towards the trans-side of the membrane.

\begin{flushleft}
{\bf {Translocation Through a Conformationally Changing Pore}}
\end{flushleft}

A more realistic description of the translocation can be obtained
by taking into consideration fluctuations in the {\it TPP}, either
spontaneous or interaction induced. Accordingly, we introduce an
additional pore conformation which is represented by the
propagation matrix $\mathbf{B}$. The changes in the pore
conformation between $A$ and $B$ are controlled by the
interconversion rates, $\omega_A$ and $\omega_B$. $\omega_A$
($\omega_B$) is the rate of the change from the $A$ ($B$) to the
$B$ ($A$) pore conformation.

The physical picture of the process is that when the pore
conformation changes, a different environment is created for the
ssDNA occupying the {\it TPP}. This implies a change in $\xi_p$
and $\mu$. For a large polymer, $N>d$, we take
$\mathbf{B}\approx\lambda \mathbf{A}$, where $\lambda$ is a
(dimensionless) parameter that represents the effect of the
conformational change on $\xi_p$ and $\mu$. The parameter
$\lambda$ may be interpreted as a measure of an effective
available volume in the {\it TPP}, when the amino acids residues
protruding the {\it TPP} change their positions.

The equations of motion of the ssDNA translocation through the
fluctuating pore, written in matrix representation, are:
\begin{equation}\label{6}
   \frac{\partial}{\partial t}
       \begin{array}{cc}
       \left( \begin{array}{c}
       \overrightarrow{P}(t;A) \\
       \overrightarrow{ P}(t;B) \\
       \end{array}\right)
        =
       \left( \begin{array}{cc}
       \mathbf{A}-\mathbf{\omega_A} & \mathbf{\omega_B} \\
       \mathbf{\omega_A} & \mathbf{B}-\mathbf{\omega_B} \\
       \end{array}\right)
       \left(\begin{array}{c}
       \overrightarrow{P}(t;A) \\
       \overrightarrow{P}(t;B) \\
       \end{array}\right),
  \end{array}
 \end{equation}
where $\overrightarrow{P}(t;i), i=A,B$ is the occupation {\it pdf}
vector of configuration $i$,
$\mathbf{\omega_i}=\omega_i\mathbf{I}$, and $\mathbf{I}$ is the
unit matrix of $n$ dimensions.

As a general note we refer to the form of Eq. 6, which was used to
study the resonant activation phenomenon \cite{barhaimb}. This
phenomenon, which was first reported by Doering and Gadoua (1992),
is the occurrence of a global minimum in the MFPT as a function of
the interconversion rate for a system in which
$\omega_A=\omega_B$. Because of the assumption $\mathbf{B}=\lambda
\mathbf{A}$, the system investigated here cannot exhibit this
phenomenon \cite{floma}.

\begin{flushleft}
{\bf DENSITY OF TRANSLOCATION TIMES}
\end{flushleft}
\begin{flushleft}
{\bf {Parameters Tuning}}
\end{flushleft}
To study the translocation of ssDNA through a fluctuating pore, we
start by computing $F(t)$. Formally, $F(t)$ is defined by
\begin{equation}\label{7}
F(t)=\partial(1-S(t))/\partial t.
\end{equation}
Here, $S(t)$ is the survival probability, namely, the probability
to still have at least one monomer in the pore, and is given by
summing the elements of the vector that solves Eq. 6; see Appendix
A for details. Using the known values of $\xi_p$ and $\mu$ from
the single conformation model, we examine in this subsection the
effect of the parameters $\lambda$, $\omega_A$ and $\omega_B$ on
$F(t)$.

First, we check the effect of $\lambda$ on $F(t)$ for several
limiting cases. For $\lambda$=$0$ movement in any direction occurs
only under the $A$ conformation environment. The $B$ conformation
traps the polymer for a period of time governed by the
interconversion rates. For $\lambda$=$1$, namely,
$\mathbf{B}$=$\mathbf{A}$, the environmental changes do not affect
the translocation, and the process reduces to a translocation
through a single conformation. For $\lambda>1$ the environmental
changes enhance the process. In this paper we restrict ourselves
to the range $0\leq\lambda\leq 1$.
\begin{figure}[b]
\includegraphics[width=1.00\linewidth,angle=0]{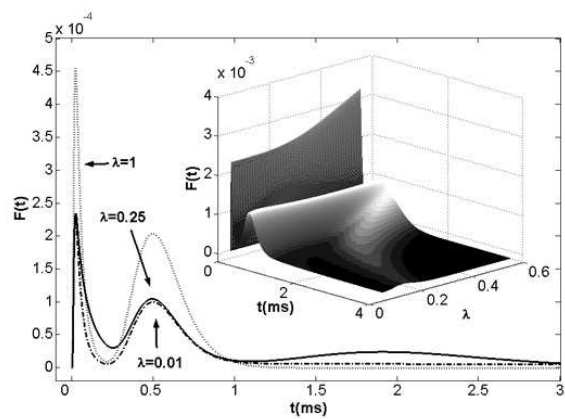}
\caption{ Poly-$dT_{nu}$ $F(t)$, for several values of $\lambda$,
with: $N$=$30$, $d$=$12$, $x$=$N+d/2$, $T$=$2^oC$, $V/V_C$=$2$,
$\omega_B$=$10^{2}$Hz, $\omega$=$1$, and $z\approx 1/2$. The left
peak represents the non-translocated events, whereas the other two
peaks represent translocation. Inset: The range for which
$\lambda$ yields three peaked $F(t)$ is shown to be
$0.1\leq\lambda\leq0.3$, when given the above parameters.}
\end{figure}

The picture is less intuitive for intermediate values of
$\lambda$. Fig. 1 shows that by choosing $\lambda$ properly, three
peaks in $F(t)$ can be obtained. In particular, as shown in the
inset of Fig. 1, the range of $\lambda$ values for which $F(t)$
exhibits three distinct peaks is $0.1\leq\lambda\leq0.30$. For the
single conformation case we found that $F(t)$ can be either mono
or double peaked depending on $V/V_C$, and on the initial state of
the translocation $x$. The short time peak represents the non
translocated events, while the long time peak represents the
translocation events. The generalization for two pore
conformations may yield two translocation peaks in addition to a
short time non translocation peak. Indeed, Fig. 1 supports the
expected behavior for the limiting $\lambda$ values, and shows
that as $\lambda\rightarrow 1$, $F(t)$ possess only one
translocation peak, as well as for $\lambda\rightarrow 0$, where
the $B$ conformation peak spreads out towards larger times, which
results in its disappearance.

Although Fig. 1 is obtained for a given value of the
interconversion rates, our explanations regarding the $F(t)$
behavior for the limiting cases $\lambda$=$1,0$, are valid for any
system conditions. This is demonstrated by calculating the MFPT
(Appendices B and D). In Appendix B we show that when
$\lambda$=$1$, the MFPT of the two configurations model reduces to
that of the single conformation model. In Appendix D we show that
for $\lambda$=$0$, the $B$ conformation contribution for the MFPT
is a term which is inversely proportional to the interconversion
rate, $\omega^{-1}_B$.

Therefore, $\lambda$ serves as a tuning parameter that leads to
either one or two actual translocation peaks in $F(t)$. The
question of interest is how $\lambda$ depends on the system
parameters. We assume a small field perturbation in the regime of
biological interest [$0\leq V/V_C\leq3$, using $V_C\approx 50 mV$
\cite{flom}], so that $\lambda(V)$ follows
$\lambda\approx\lambda_0+V/V_{\lambda}$, and keeping
$\lambda(V)\leq 1$. Here $\lambda_0$ and $V_{\lambda}$ might be
expansion coefficients, where $\lambda_0\ll 1$ is implied from
recent experiments \cite{meller2}, as we discuss later. The
process can be viewed such that as the voltage increases, amino
acids residues protruding the {\it TPP} that constitute obstacles
for the translocating ssDNA clear the way. While the $\lambda$
dependence on the voltage is assumed here, its dependence on other
system parameters (e. g., temperature and pH) is unknown and is
folded into $V_{\lambda}$.

To check how interconversion rates affect $F(t)$, it is convenient
to define two dimensionless parameters,
$\omega\equiv\omega_A/\omega_B$ and $\omega_B/k$ (or
$\omega_A/k$), where $k$ is the dominate rate of the $A$
conformation for a sufficiently large $N$, $k=R/d^{\mu}$. The
first ratio set the dominance of a given conformation over its
counterpart; namely, for $\omega\ll 1$ most of the translocation
events take place in the $A$ conformation. The second ratio gives
an estimate of the number of moves done in a given conformation
before a change in the pore structure occurs, and thus relates the
ssDNA dynamics to the structural changes dynamics.

As shown in Fig. 2 and in the inset of Fig. 2, $F(t)$ exhibits two
peaks corresponding to actual translocation only when
$\omega\approx1$. For $\omega\ll 1$ and $\omega\gg 1$ only one
peak corresponding to an actual translocation survives. For all
cases there is a peak representing non-translocation events. In
addition, we find that for two translocation peaks to be obtained,
the ratio $\omega_B/k$ (or $\omega_A/k$ due to $\omega\approx 1$)
must fulfil $\omega_B/k\leq10^{-3}$ (data not shown). The lower
limit of the interconversion rates is inversely proportional to
the order of the measurement time otherwise only one conformation
will be detected.

Finally, we assume that the rate of the conformational changes is
controlled mainly by temperature; namely, we take $\omega_A$ and
$\omega_B$ as voltage independent in the regime of biological
interest: $0\leq V/V_C\leq3$.

\begin{figure}[t]
\includegraphics[width=1.00\linewidth,angle=0]{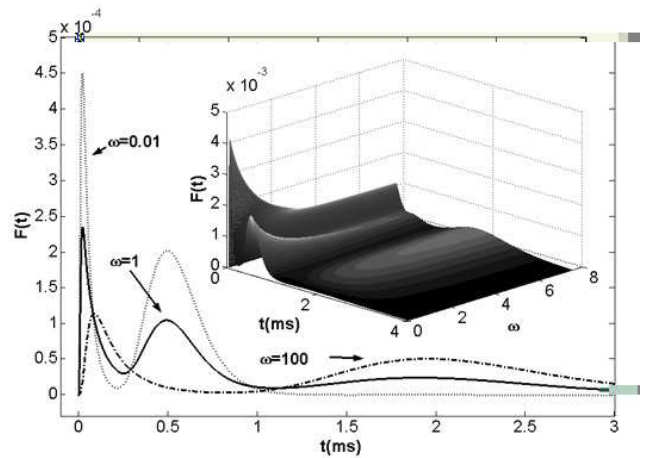}
\caption{Poly-$dT_{nu}$ $F(t)$, for several values of $\omega_A$
and fixed $\omega_B$ ($\omega_B$=$10^2$Hz), with $\lambda=1/4$,
and the other parameters as in Fig. 1. Inset: For small values of
$\omega$, $\omega\lesssim 10^{-2}$, $F(t)$ displays one
translocation peak that corresponds to conformation $A$, whereas
for large values of $\omega$, $\omega\gtrsim 10^{2}$, $F(t)$
displays one translocation peak that corresponds to conformation
$B$. For $\omega\approx 1$ two translocation peaks are obtained.}
\end{figure}

\begin{flushleft}
{\bf {Translocation Velocity}}
\end{flushleft}

To study further the translocation process, we check the voltage
dependence of the times that maximize the peaks of $F(t)$, denoted
as $t_{m,i}$ where $i$=$1,2,3$ (e.g. $t_{m,1}$ characterized the
short time peak). In previous works \cite{flom,meller1},
$t^{-1}_{m}$ for one translocation peak was regarded as the most
probable average velocity of the translocation (up to a
multiplicative constant). We show below that our assumptions
regarding the voltage dependence of the system parameters, yield
either linear or quadratic dependence of the translocation
velocity on the voltage, and can be used to explain the different
experimental observations.

Fig. 3 shows  $t_{m,i}(V_C/V)$ and $t_{m,i}^{-1}(V/V_C)$, for
$V_{\lambda}$=$350mV$, in a voltage window that leads to
$0.215\leq\lambda(V)\leq 0.30$, and accordingly to a triple-peaked
$F(t)$. $t_{m,1}$ is almost independent of $V_C/V$ (see Fig. 3a).
Although the non translocation peak amplitude decreases upon
increasing $V/V_C$, the location of its maximum hardly changes.
This happens since exiting against the field must happen within a
short time window at the beginning of the process, otherwise the
polymer is more likely to cross the membrane due to the electric
bias. Similar behavior was observed experimentally \cite{kasa1}.

For the single conformation case, we showed that $t_{m,2}^{-1}$
scales linearly with $V/V_C$ when the initial state of the
translocation is near the cis-side of the membrane \cite{flom}.
Fig. 3b shows that the linear scaling of $t_{m,2}^{-1}(V/V_C)$
persists. However, $t_{m,3}^{-1}(V/V_C)$ (Fig. 3c) displays a
quadratic behavior, which is a consequence of the form of
$\lambda(V)$, as discuss in the next section when computing the
MFPT. On the other hand, setting $V_{\lambda}$=$120mV$ leads to
one translocation peak, and to small deviations from linearity
towards a weak quadratic behavior of $t_{m,2}^{-1}(V/V_C)$; see
Fig. 4.

The model of two conformations not only yields one or two actual
translocation peaks as a function of $V_{\lambda}$, but can
account for either a linear or quadratic dependence of the
translocation velocity with the voltage, again, as a function of
$V_{\lambda}$. Thus, varying $V_{\lambda}$ we obtain different
behaviors of the translocation, which can be related to the
different experimental observations.

\begin{figure}[t]
\includegraphics[width=1.00\linewidth,angle=0]{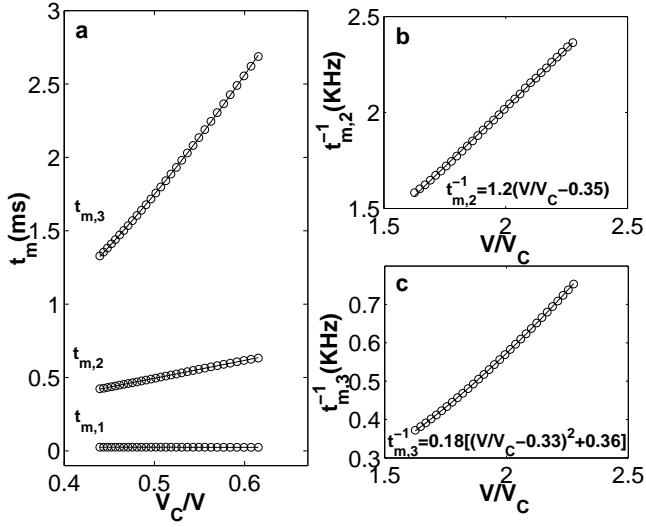}
\caption{{\bf a}: $t_{m,i}$ for poly-$dT_{nu}$, as a function of
$V/V_C$ for the same parameters as in Fig. 1 and
$V_{\lambda}=350mV$. $t_{m,1}$ is almost independent of $V_C/V$ in
contrast to the pronounced dependence of $t_{m,2}$ and $t_{m,3}$.
{\bf b-c}: $t_{m,2}^{-1}$ and $t_{m,3}^{-1}$ depend linearly and
quadratically on $V/V_C$, respectively. The solid lines through
the circles are polynomial fits.}
\end{figure}
\begin{figure}[t]
\includegraphics[width=1.00\linewidth,angle=0]{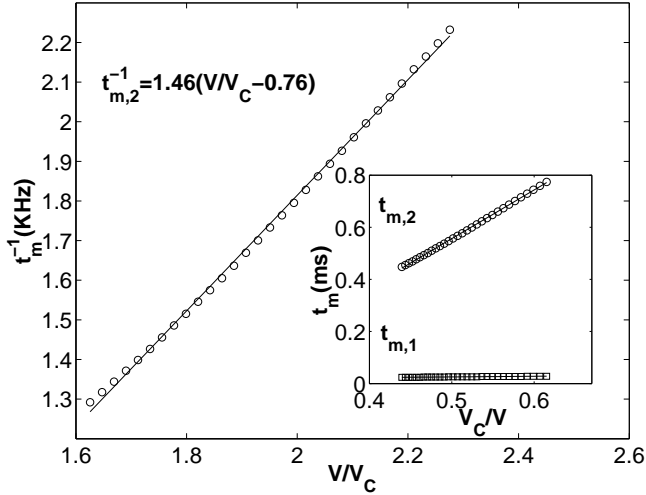}
\caption{$t_{m,2}^{-1}$ for poly-$dT_{mu}$ for the same parameters
as in Fig. 3 except for $V_{\lambda}=120mV$. This value for
$V_{\lambda}$ leads to $0.625\leq\lambda(V)\leq 0.875$ and
accordingly for one translocation peak. The solid line is a
polynomial fit. Inset: $t_{m,1}$ and $t_{m,2}$ behave
qualitatively the same as for the case $V_{\lambda}=350mV$.}
\end{figure}

\begin{flushleft}
{\bf THE MFPT}
\end{flushleft}
\begin{flushleft}
{\bf Small Field Biased Translocation}
\end{flushleft}

We now turn to calculate the MFPT, which allows for an analytical
estimation of the characteristic times of the FPT {\it pdf} and
{\it cdf}. In general, the $m$ moment of $F(t)$ is calculated by
raising to the $m$ power the inverse of the propagation matrix.
For the two conformations translocation this matrix is given on
the right hand side of Eq. 6.

After somewhat lengthly calculations, which are given is
Appendices B and C, the expression for the MFPT, $<\tau>$, reads:
\begin{equation}\label{8}
    <\tau>\approx
  \frac{\overline{\tau}}{\lambda}[(\lambda
  P_{A,0}+P_{B,0})+\overline{\tau}(\omega_A+\omega_B)/2],
\end{equation}
where $\overline{\tau}$ is the MFPT for the single configuration
model, and is given by Eq. C5, and $P_{A,0}(P_{B,0})$ is the
probability that the process starts in conformation $A$ ($B$). For
$P_{A,0}$ and $P_{B,0}$ the equilibrium condition is assumed,
$P_{A,0}=\omega_B/(\omega_A+\omega_B)$ and $P_{B,0}=1-P_{A,0}$.

Eq. 8 is valid for not too high fields, $V/V_C\gtrsim 1$, and the
relations between the interconversion with $k$ found in the
previous section, $\omega_A/k,\omega_b/k\ll 1$. The first term in
the brackets of Eq. 8, $\lambda  P_{A,0}+P_{B,0}$, represents the
translocation peaks and can be compared with $t_{m,2}$ and
$t_{m,3}$. The second term in the brackets,
$\overline{\tau}(\omega_A+\omega_B)/2$, represents the coupling
time cost, and is of the order of $o(10^{-2})$ for voltages that
obey $V/V_C\geq1.5$. Keeping the first term in Eq. 8, we have
\begin{equation}\label{9}
    <\tau>\approx
  \frac{2x\xi_pb^2d^\mu}{z|q|(1+1/d)}\frac{1}{V-V_C}(P_{A,0}+P_{B,0}\frac{V_\lambda}{V}),
\end{equation}
where $x\approx N$ for a translocation process that starts near
the cis-side of the membrane.

Eq. 9 provides a solid basis for the numerically obtained
dependence of the translocation velocity on the voltage. $<\tau>$
consists of two terms that can be attributed to the $A$ (first
term in the brackets) and $B$ (second term in the brackets) pore
conformations. For $V_{\lambda}\approx 120mV$ we can replace the
expression in brackets by one in the relevant voltages window.
Thus, we find that $<\tau>\propto (V-V_C)^{-1}$, which implies
that $F(t)$ has one translocation peak for this choice of
$V_{\lambda}$. For higher values of $V_{\lambda}$ and voltages of
biological interest, the two terms in the brackets are separated.
This leads to a term that represent the $A$ conformation and
scales as $(V-V_C)^{-1}$, and a term that represents the $B$
conformation that scales as $[V(V-V_C)]^{-1}$.

Accordingly, Eq. 9 captures the physical essence of the
translocation of the ssDNA through the conformationally changing
pore, under a relatively small field.
\\
\\
\begin{flushleft}
{\bf Field-Free Translocation}
\end{flushleft}

In recent field-free experiments by Bates et al. (2003), the {\it
cdf} $G(t)$=$\int_0^tF(s)ds$ was shown to have two regimes that
were approximated by a fast and a slow timescales, $\tau_1$ and
$\tau_2$, with the ratio $\tau_1/\tau_2\approx1/20$.
\begin{figure}[b]
\includegraphics[width=1.00\linewidth,angle=0]{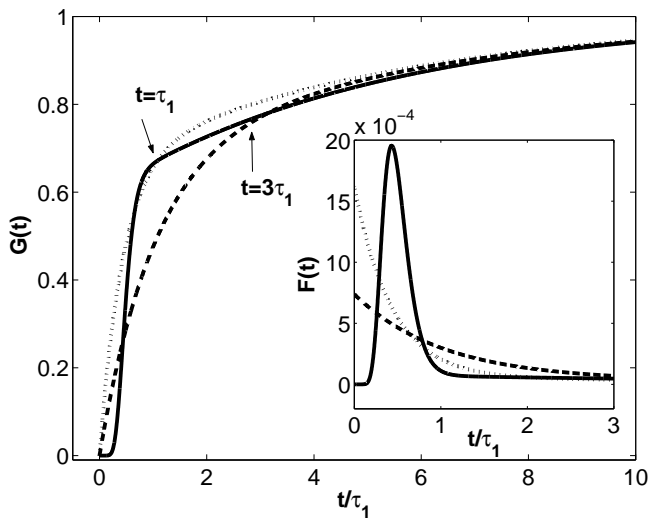}
\caption{$G(t)$ for poly-$dA_{nu}$, full curve, for $V$=$0$, and
the initial state $x$=$n/2$, $\omega_A$=$10^{-2}$Hz,
$\omega$=$1/2$ and the other parameters as in Fig. 1. Also shown
is the approximate {\it cdf} $G_{ap}(t)$, dashed curve, and its
modified version, dotted curve. Inset: $F(t)$ for poly-$dA_{nu}$
for the corresponding {\it cdf} shown in the main Fig.}
\end{figure}
Motivated by these experimental result, which implies ,within our
approach, that $\lambda_0$ fulfils $\lambda_0\ll 1$, we study in
this subsection the zero field translocation, $V\rightarrow 0$. We
start by computing $G(t)$ for a translocation process that starts
at the middle state, $x=n/2$. This is the the same initial
condition that was imposed in the experiments \cite{meller2}. As
shown in Fig. 5 (full curve), $G(t)$ displays two regimes, a fast
increase at short times and a slow increase from intermediate to
large times. Accordingly, we try the approximation
\begin{equation}\label{10}
 G_{ap}(t)\approx 1-(P_{A,0}e^{-t/\tau_1}+P_{B,0}e^{-t/\tau_2}).
\end{equation}
Matching the first and the second moments obtained from $F(t)$ and
from the approximated $F(t)$ we find that the characteristics
timescales of $G_{ap}(t)$ are (see Appendix D):
\begin{equation}\label{11}
    \tau_1=\overline{\tau}(1+3\omega/2);~~~~~\tau_2=\overline{\tau}(1/2+\omega)+1/\omega_B,
\end{equation}
which, when used in $G_{ap}(t)$, lead to the dashed curve plotted
in Fig. 5. Also shown, dotted curve, is a modified version of
$G_{ap}(t)$, where $\tau_1\rightarrow t_m$ is used in Eq. 10. Note
that for the short times, $t<\tau_1$, the later approximation fits
better $G(t)$, but from intermediate times, $t>3\tau_1$,
$G_{ap}(t)$ and $G(t)$ coincide.

The two conformations model produces a temporal behavior that
agrees with experimental observation, and provides a good
explanation for it. In the limit, $V\rightarrow0$, the $B$
conformation acts as a trapping conformation; namely, the polymer
is stuck in its position when subject to the environment due to
the $B$ conformation. Movement occurs only through the $A$
conformation. As a result two regimes are obtained for $G(t)$. The
fast increase in $G(t)$ at short times, is a consequence of
exiting due to the $A$ conformation (at either side of the
membrane), while the slow saturation at longer times is a result
of the release from the trapping in the $B$ conformation.

In the inset of Fig. 5 we show both $F(t)$, the approximate
$F(t)$, and the modified version of the approximation, which is
obtained when using $\tau_1\rightarrow t_m$. Because the process
starts in the middle state, $x$=$n/2$, $F(t)$ has only one peak,
which coincides with previous results \cite{flom}. Although the
approximate $F(t)$, or any other approximation of two exponentials
with positive coefficients, does not exhibit a peaked shape,
information about the maximal peak value of $F(t)$ and the
interconversion rates can still be extracted from $G_{ap}(t)$
timescales by using Eq. 11. For example, the timescales suggested
by Bates et al. (2003), imply that $t_m\approx 165\mu s$, and
$\omega_B\approx 300$Hz.

\begin{flushleft}
{\bf CONCLUSIONS}
\end{flushleft}

The model introduced here describes the translocation of ssDNA
through a fluctuating pore structure. As a consequence the ssDNA
within the transmembrane pore part is exposed to a changing
environment, which can be reflected in the first passage times
{\it pdf}, $F(t)$. By computing $F(t)$, comparing our results to
experimental results, and using physical arguments, we obtained
theoretically a behavior which was previously observed
experimentally - $F(t)$ having three peaks. This behavior is
obtained by tuning the dimensionless parameter $\lambda$, which
controls the effect of the change in the pore structure on the
translocating ssDNA, and the interconversion rates between the
pore conformations, $\omega_A$ and $\omega_B$. In particular,
$\lambda$ has to fulfill $0.1\leq\lambda\leq0.30$, and the
interconversions rates have to be of the same order of magnitude,
and much smaller than the typical rate of the $A$ pore
conformation, $k$, $\omega_B/k\leq 10^{-3}$. This implies that the
relaxation timescale of the ssDNA in the pore is much shorter than
of the pore conformational changes timescale. From these relations
the maximal values of the interconversion rates can be deduced
from the value of $k$, given by Eq. 3, to be:
$\omega_A\approx\omega_B$=$10^2$Hz.

We have been able to show, both numerically and analytically, that
the times that maximize the $F(t)$ actual translocation peaks,
$t_{m,i}, i=2,3$, and the MFPT, are inversely proportional to the
first or the second power of the field, depending on
$V_{\lambda}$. This emphasizes the crucial role of $V_{\lambda}$
on the translocation extracted functions, and may explain the
different experimental results for $F(t)$ discussed in the
introduction, meaning that $V_{\lambda}$ is sensitive for the
specific experimental set up, and biological conditions.

The probability to exit the channel until time $t$, $G(t)$, in a
field-free environment, has been shown to have two regimes that
can be approximated by two timescales, $\tau_1$ and $\tau_2$,
which are about one order of magnitude apart, and are closely
related to the $\overline{\tau}$, $t_{m}$, and the interconversion
rates: $\tau_1=\overline{\tau}(1+3/2\omega)$, or
$\tau_1\rightarrow t_m$, and
$\tau_2=\overline{\tau}(1/2+\omega)+1/\omega_B$. From these
relations the interconversion rates can be deduced when analyzing
experimental data.

\acknowledgments we acknowledge fruitful discussions with Ralf
Metzler and with Amit Meller, and the support of the US-Israel
Binational Science Foundation and the Tel Aviv University
Nanotechnology Center.

\begin{flushleft}
{\bf APPENDIX A}
\end{flushleft}

In this appendix we introduce the formal solution of Eq. 6 and
define the symbols used in next derivations. In general, $S(t)$
for a discrete system is given by summing the elements of the
vector that solves the ME,
\renewcommand{\theequation}{A1}
\begin{equation}\label{A1}
S(t)=\overrightarrow{U}_{2n}\mathbf{E}e^{\mathbf{D}t}\mathbf{E}^{-1}\overrightarrow{P}(0|2n).
\end{equation}
Here $\overrightarrow{U}_{2n}$ is the summation row vector of $2n$
dimensions, $\overrightarrow{P}(0|2n)$ is the initial condition
column vector,
\renewcommand{\theequation}{A2}
\begin{equation}\label{A2}
[\overrightarrow{P}(0|2n)]_j=(P_{A,0}\delta_{x,j}+P_{B,0}\delta_{x+n,j}),
\end{equation}
where $x$ is the initial state of the translocation process. The
definite negative real part eigenvalues matrix, $\mathbf{D}$, is
obtained through the similarity transformation:
$\mathbf{D}=\mathbf{E}^{-1}\mathbf{H}\mathbf{E}$, where
$\mathbf{H}$ is the matrix given on the right hand side of Eq. 6,
and $\mathbf{E}$ and $\mathbf{E}^{-1}$ are the eigenvectors
matrix, and its inverse, of $\mathbf{H}$.

\begin{flushleft}
{\bf APPENDIX B}
\end{flushleft}

Here we calculate formally the MFPT ~$<\tau>$. The $m$ moment of
$F(t)$ is given by: ${<\tau^m>}=\int _0^\infty t^mF(t)dt=
m!\overrightarrow{U}_{2n}(-\mathbf{H})^{-m}\overrightarrow{P}(0|2n)$.
To calculate the inverse of the propagation matrix $\mathbf{H}$,
which is given on the right hand side of Eq. 6, we use the
projection operator \cite{kla,zwa}:
$\mathbf{Q}\mathbf{H}\mathbf{Q}\equiv
\mathbf{H}_{\mathbf{QQ}}=\mathbf{A}-\mathbf{\omega_{A}}$,~
$\mathbf{H}_{\mathbf{QZ}}=\mathbf{\omega_{B}}$,~
$\mathbf{H}_{\mathbf{ZQ}}=\mathbf{\omega_{A}}$,~
$\mathbf{H}_{\mathbf{ZZ}}=\mathbf{B}-\mathbf{\omega_{B}}$,~and the
identity, $\mathbf{I}=\mathbf{H}\mathbf{M}$, and obtain
$\mathbf{M}$ blocks:
\renewcommand{\theequation}{B1}
\begin{eqnarray}
 \mathbf{M}_{\mathbf{QQ}}=[\mathbf{A_{QQ}}-\mathbf{A_{QZ}(A_{ZZ})}^{-1}\mathbf{A_{ZQ}}]^{-1}\nonumber \\
 =\mathbf{A}^{-1}\mathbf{C}(\mathbf{B}-\mathbf{\omega_B})~~~~~~\nonumber \\
 \mathbf{M}_{\mathbf{QZ}}=[\mathbf{A_{ZQ}}-\mathbf{A_{ZZ}(A_{QZ})}^{-1}\mathbf{A_{QQ}}]^{-1}\nonumber \\
 =-\mathbf{A}^{-1}\mathbf{C}\mathbf{\omega_B},~~~~~
\end{eqnarray}
where $\mathbf{M_{ZQ}}$ and $\mathbf{M_{ZZ}}$ are obtained in a
similar way. Now, we can write the $m$ moment vector of $F(t)$ as
\renewcommand{\theequation}{B2}
\begin{equation}\label{B2}
  \begin{array}{cc}
    \begin{array}{c}

     <\overrightarrow{\tau^m}> \\
    \end{array}
    =
    m{!} \left(-\mathbf{M} \right)^m
    \begin{array}{c}

      \overrightarrow{P}(0|2n) \\
\end{array}
  \end{array},
\end{equation}
where $\mathbf{M}$ is given by
\renewcommand{\theequation}{B3}
\begin{equation}\label{B3}
\mathbf{M}=\left( \begin{array}{cc}
      \mathbf{A}^{-1}\mathbf{C}(\mathbf{B}-\mathbf{\omega_B}) & -\mathbf{A}^{-1}\mathbf{C}\mathbf{\omega_B} \\
      -\mathbf{A}^{-1}\mathbf{C}\mathbf{\omega_A} & \mathbf{A}^{-1}\mathbf{C}(\mathbf{A}-\mathbf{\omega_A}) \\
    \end{array}\right),
\end{equation}
and
$\mathbf{C}$=$(\mathbf{B}-\mathbf{\omega_B}-\lambda\mathbf{\omega_A})^{-1}$.
For $m=1$ in Eq. B2 we obtain the MFPT vector:
\renewcommand{\theequation}{B4}
\begin{equation}\label{B4}
    <\overrightarrow{\tau}>=\left(\begin{array}{c}
      -\mathbf{A}^{-1}\mathbf{C}(P_{A,0}\mathbf{B}-\mathbf{\omega_B}) \overrightarrow{P}(0|n)\\
      -\mathbf{A}^{-1}\mathbf{C}(P_{B,0}\mathbf{A}-\mathbf{\omega_A}) \overrightarrow{P}(0|n)\\
    \end{array}\right),
\end{equation}
where $[\overrightarrow{P}(0|n)]_j$=$\delta_{x,j}$. Summing
$<\overrightarrow{\tau}>$ elements by using the summation row
vector of $n$ dimensions $\overrightarrow{U}_n$, results in:
\renewcommand{\theequation}{B5}
\begin{eqnarray}\label{B5}
    <\tau>=-\overrightarrow{U}_n\mathbf{C}\overrightarrow{P}(0|n)(\lambda
    P_{A,0}+P_{B,0})+ \nonumber \\
    +\overrightarrow{U}_n\mathbf{A}^{-1}\mathbf{C}\overrightarrow{P}(0|n)(\omega_A+\omega_B).
\end{eqnarray}
Note that the MFPT of the single $A$ conformation,
$\overline{\tau}$, is:
$\overline{\tau}$=$-\overrightarrow{U}_{n}\mathbf{A}^{-1}\overrightarrow{P}(0|n)$,
which has a similar form as of the first term in Eq. B5 when
choosing $\mathbf{C}^{-1}$ as the propagation matrix.

It is easy to verify that for $\lambda$=1, $<\tau>$ reduces to the
MFPT of the single conformation case, $\overline{\tau}$. Rewriting
Eq. B5 as
\renewcommand{\theequation}{B6}
\begin{eqnarray}\label{B6}
    <\tau>=-\overrightarrow{U}_n\mathbf{A}^{-1}\mathbf{C}[P_{A,0}\mathbf{B}+P_{B,0}\mathbf{A}- \nonumber \\
    -\mathbf{\omega_A}-\mathbf{\omega_B}]\overrightarrow{P}(0|n),~~~~~~~~~~
      \end{eqnarray}
and substituting $\lambda$=$1$ we find that
\renewcommand{\theequation}{B7}
\begin{equation}\label{B7}
 <\tau>=-\overrightarrow{U}_n\mathbf{A}^{-1}\overrightarrow{P}(0|n)=\overline{\tau}.
\end{equation}

\begin{flushleft}
{\bf APPENDIX C}
\end{flushleft}

In order to obtain an explicit expression for the MFPT of the
translocation in a weak field limit, we first rewrite Eq. B5 as
\renewcommand{\theequation}{C1}
\begin{eqnarray}\label{C1}
     <\tau>= \widehat{\tau}(\lambda P_{A,0}+P_{B,0})+\widetilde{\sigma^2}(\omega_A+\omega_B),
\end{eqnarray}
where
$\widehat{\tau}=-\overrightarrow{U}_n\mathbf{C}\overrightarrow{P}(0|n)$,
and
$\widetilde{\sigma^2}=\overrightarrow{U}_n\mathbf{A}^{-1}\mathbf{C}\overrightarrow{P}(0|n)$.
We can further rewrite $\widehat{\tau}$ as
$\widehat{\tau}=\sum_{s=1}^n\widehat{\tau}_{s,x}$.
$\widehat{\tau}_{s,x}\equiv -(\mathbf{C})_{s,x}$ defines the mean
residence time (MRT) spent in state $s$ before exiting the
channel, given that the process started at state $x$
\cite{barhaima}, and has the form \cite{huang}:
\renewcommand{\theequation}{C2}
\begin{equation}\label{C2}
  -(\mathbf{C})_{s,x}=\frac{\Delta (h^s)\Delta (h^{n+1-x})}{\Delta (h)\Delta
  (h^{n+1})}\frac{r^{x-s}}{\widehat{k}};~~~~~~~~~s<x,
\end{equation}
where $(\mathbf{C})_{s,x}$ for $s\geq x$, is obtained when
exchanging $x$ for $s$ and $r$ for $l$ in Eq. C2. Here
$h_{\pm}$=$[1\pm(1-4rl)^{1/2}]/2$, $r$=$ap_+$, $l$=$ap_-$,
$a$=$\lambda k/\widehat{k}$, and $\widehat{k}$=$\lambda
k+\omega_B+\lambda\omega_A$. Thus, we find that
$\widehat{\tau}_{s,x}$ is a function of the parameter
$a=[1+(\omega_A+\omega_B/\lambda)/k]^{-1}$, which obeys $0\leq
a\leq 1$, and is a measure of the difference between
$\overline{\tau}$ and $\widehat{\tau}$. Using $\omega\approx 1$
and $\omega_A/k\approx 10^{-3}$ leads to $a\approx 1$ given
$V/V_C\geq 1$, and accordingly to
\renewcommand{\theequation}{C3}
\begin{equation}\label{C3}
  (\mathbf{C})_{s,x}=(\mathbf{A}^{-1})_{s,x}k/\widehat{k}=(\mathbf{A}^{-1})_{s,x}/\lambda.
\end{equation}
Eq. C3 implies
\renewcommand{\theequation}{C4}
\begin{equation}\label{C4}
  \widehat{\tau}=\overline{\tau}/\lambda.
\end{equation}
To obtain an explicit expression for $\overline{\tau}$, it is
convenient to use the independence approximation and replace
$p_{j,j-1}$ and $k_j$ by state independent terms:
$p_+=[1+e^{(-V/V_C+1)}]^{-1}$ and $k$. This approximation, which
is valid for large polymers, $N>d$, and becomes more accurate as
$N$ increases, leads to $a_+=p_+k$, $a_-=(1-p_+)k$, so that
\cite{flom}:
\renewcommand{\theequation}{C5}
\begin{equation}\label{C5}
\overline{\tau} = \frac{\Delta (p^{n+1-x})p_+^xx-\Delta
(p^{x})p_-^{n+1-x}(n+1-x)}{k\Delta p \Delta (p^{n+1})},
\end{equation}
where $\Delta (p^s)=p_+^s-p_-^s$. In the limit of a not too large
field, $V/V_C\gtrsim 1$, Eq. C5 reduces to
\renewcommand{\theequation}{C6}
\begin{equation}\label{C6}
  \overline{\tau}\approx
  \frac{2x\xi_pb^2d^\mu}{z|q|(1+1/d)}\frac{1}{V-V_C}.
\end{equation}
To compute $\widetilde{\sigma^2}$ we rewrite
$\widetilde{\sigma^2}$ as
$\widetilde{\sigma^2}=\Sigma_{s=1}^{n}\overline{\tau}_s\widehat{\tau}_{s,x}$,
where $\overline{\tau}_s$ is given by Eq. C5 for $x=s$, and
$\widehat{\tau}_{s,x}$ is given by Eq. C2. For $a\approx 1$ we
have $\widetilde{\sigma^2}=\overline{\tau^2}/2\lambda$, where
$\overline{\tau^2}$ is the second moment of $F(t)$ for the single
$A$ conformation case. The calculation of
$\overline{\tau^2}/2=\sum_{s=1}^n\overline{\tau}_s\overline{\tau}_{s,x}$
yields in the weak field limit $V/V_C\gtrsim 1$:
\renewcommand{\theequation}{C7}
\begin{equation}\label{C7}
    \frac{\overline{\tau^2}}{2}\approx \frac{1}{(k\Delta
    p)^2}[\frac{x(x-1)}{2}+\frac{xy^x(1-y)-y(1-y^x)}{(1-y)^2}+II],
\end{equation}
where $y=p_-/p_+$ and
\renewcommand{\theequation}{C8}
\begin{eqnarray}\label{C8}
    II= \frac{y^{-x}-1}{y^{-n}-1}\left[\frac{p_-^{x-n-1}}{1/p_--1}\right(\frac{p_+^{n+1-x}-1}{1-p_+}+n+1-\nonumber~~~~~~~~~~~~~~~~~\\
    -xp_+^{n+1-x}\left)-(n+1-x)\frac{n+x}{2}\right].~~~~(C8)\nonumber~~~~~~~~
     \end{eqnarray}
Noticing that $II$ represents the non-translocation events and
vanishes for $V/V_C\gtrsim 1$ as $y^{n-x}$, we rewrite Eq. C7 up
to a leading term in $x$ as:
\renewcommand{\theequation}{C9}
\begin{equation}\label{C9}
    \frac{\overline{\tau^2}}{2}\approx \frac{x(x-1)}{2(k\Delta
    p)^2}.
\end{equation}
Using $\overline{\tau}\approx (x/k\Delta p)$ valid for
$V/V_C\gtrsim 1$ \cite{flom}, Eq. C9 yields for a leading order in
$x$
\renewcommand{\theequation}{C10}
\begin{equation}\label{C10}
   \widetilde{\sigma^2}\approx
\overline{\tau}^2/2\lambda.
\end{equation}
Substituting Eq. C4 and Eq. C10 into Eq. C1, Eq. 8 is obtained.
\\
\\
\begin{flushleft}
{\bf APPENDIX D}
\end{flushleft}

For the analysis of the field-free translocation we start by
computing $<\tau>$ and $<\tau^2>$. Substitute $\lambda$=$0$ in Eq.
B5, we obtain
\renewcommand{\theequation}{D1}
\begin{equation}\label{D1}
    <\tau>=-\overrightarrow{U}_n\mathbf{A}^{-1}\overrightarrow{P}_n(0)\left(\frac{\omega_A+\omega_B}{\omega_B}\right)+\frac{P_{B,0}}{\omega_B},
\end{equation}
which can be written as
\renewcommand{\theequation}{D2}
\begin{equation}\label{D2}
  <\tau>=\overline{\tau}(1+\omega)+P_{B,0}/\omega_B.
\end{equation}
Note that experiments suggest that $\lambda_0\ll 1$, which leads
to $\lambda_0k<\omega_B$, whereas $\lambda_0k\ll\omega_B$ is used
for simplification, and enables the substitution of $\lambda$=$0$
in Eq. B5.

To compute~$<\tau^2>$,~we have to calculate the blocks of
$\mathbf{M}^2$:
\renewcommand{\theequation}{D3}
\begin{eqnarray}\label{D3}
    \mathbf{M_{QQ}}^2=\mathbf{A}^{-2}(1+\omega);~~~~~~~~~~~~~~~~~~~~~~~~~~~~~~~~~~~~~~~\nonumber\\
    \mathbf{M_{QZ}}^2=\mathbf{A}^{-2}(1+\omega)-\mathbf{A}^{-1}/\omega_B;~~~~~~~~~~~~~~~~~~~~~~~~\nonumber\\
    \mathbf{M_{ZQ}}^2=\mathbf{A}^{-2}\omega(1+\omega);\nonumber~~~~~~~~~~~~~~~~~~~~~~~~~~~~~~~~~~~~~\\
    \mathbf{M_{ZZ}}^2=\mathbf{A}^{-2}\omega(1+\omega)-\mathbf{A}^{-1}2\omega/\omega_B+1/\omega_B^2.~~~~~~~~
\end{eqnarray}
Substituting Eq. D3 into Eq. B2 and summing the vector elements,
we obtain
\renewcommand{\theequation}{D4}
\begin{equation}\label{D4}
  \frac{<\tau^2>}{2}=
  \frac{\overline{\tau^2}}{2}(1+\omega)^2+
  \overline{\tau}(1+2\omega)\frac{P_{B,0}}{\omega_B}+
  \frac{P_{B,0}}{\omega_B^2}.
\end{equation}
To get the relaxation timescales of $G_{ap}(t)$, $\tau_1$ and
$\tau_2$, we match $<\tau>$ and $<\tau^2>$ obtained from
\renewcommand{\theequation}{D5}
\begin{equation}\label{D5}
    F_{ap}(t)=\frac{P_{A,0}}{\tau_1}e^{-t/\tau_1}+\frac{P_{B,0}}{\tau_2}e^{-t/\tau_2},
\end{equation}
to the corresponding moments obtained from Eq. D2 and Eq. D4. This
procedure yields:
\renewcommand{\theequation}{D6}
\begin{equation}\label{D6}
    \tau_1=\frac{<\tau>-P_{B,0}\tau_2}{P_{A,0}}
\end{equation}
and
\renewcommand{\theequation}{D7}
\begin{equation}\label{D7}
    \tau_2=<\tau>+\left[\frac{P_{A,0}}{P_{B,0}}\left({\frac{<\tau^2>}{2}-<\tau>^2}\right)\right]^{1/2}.
\end{equation}
Substituting Eq. D2 and Eq. D4 into Eq. D7 results in
\renewcommand{\theequation}{D8}
\begin{eqnarray}\label{D8}
\tau_2=\overline{\tau}(1+\omega)+\frac{P_{B,0}}{\omega_B}+\frac{P_{A,0}}{\omega_B}\nonumber~~~~~~~~~~~~~~~~~~~~~~~~~~\\
    \left[{1-\overline{\tau}\frac{\omega_B}{P_{A,0}}+
    \frac{\omega_B^2}{P_{A,0}P_{B,0}}\left(\frac{\overline{\tau^2}}{2}-\overline{\tau}^2\right)}\right]^{1/2}.
\end{eqnarray}
Expanding the square root in Eq. D8 to leading order and using Eq.
D6, Eq. 11 is obtained.

\end{document}